\documentclass[twoside,twocolumn]{article}
\usepackage{amsmath,amssymb,wasysym}
\usepackage[OT2,T1]{fontenc}
\usepackage[utf8]{inputenc}
\usepackage[russian,frenchb,english]{babel}
\usepackage{hyperref}
\usepackage{graphicx}
\usepackage{color}
\usepackage{bm}
\usepackage{multirow}

\author{L. Jaouen$^{1)}$, F. Chevillotte$^{1)}$\\
$^{1)}$ Matelys, 7 rue des Maraîchers, Bâtiment B, 69120 Vaulx-en-Velin, France.\\
\hspace*{8pt}luc.jaouen@matelys.com, fabien.chevillotte@matelys.com}

\title{Length correction of 2D discontinuities or perforations at large wavelengths and for linear acoustics}

\date{}

\begin{document}

\maketitle

\abstract{This work focuses on the length correction due to radiation effect of a duct discontinuity or at the surface of a perforated plate in the linear acoustic domain and at large wavelengths. 
Two results are obtained from the comparison of models used in the literature for different geometries of duct discontinuities or perforations.
First, the mode-coupling at the discontinuity between two ducts or a perforated plate and air is negligible for low frequencies.
Second, all consistent models of the length correction proposed so far, for any geometries except slit-like, are similar when plotted as a function of the square root of the perforation rate (the ratio of the duct cross-section areas for ducts). These two results imply the effect of the thickness of the ducts or the perforated plate can be accounted for in the length correction expression independently from other geometric variables.
A single approximate formulation for all geometries of perforations or duct discontinuities, except slit-like geometries, is proposed for practical uses.}

\section{Introduction}

The sound radiation of a duct or perforation is a fundamental phenomenon in acoustics.
It relies on the line flow distortion around the aperture of this discontinuity or perforation. Modeling of the sound radiation and the associated line flow distortion has been a research topic widely studied since the 19th century with first attempts by e.g. G. Wertheim \cite{Wer51} or H. von Helmholtz \cite{Hel60}. A plethora of publications can be found in the literature on this subject which is of interest in musical acoustics, building acoustics or transport acoustics. Thus, this article cannot be an exhaustive review on this topic. \\

The sound radiation can be described using the concept of length correction $\varepsilon$ (or functions involving this parameter such as the "mass correction", the "inductance correction" by analogy to the electrical problem, the imaginary part of the impedance...). Indeed, the length correction, computed assuming a non-dissipative fluid, is a length addition to the geometrical duct or perforation length which defines the position of the nearest pressure node outside the duct or perforation. Note that this pressure node is virtual.
While this length correction is a frequency dependent quantity, it can be shown to be approximately constant for low frequencies, roughly defined as wavenumber in air ($k_0$) times inner radius of the smallest duct or  perforation ($r$) much smaller than 1 ($k_0 r \ll 1$, see e.g. \cite{LS48,NYI60,CKL84,NS89,DNJ01}).
With a wave velocity of  $~$340 m.s$^{-1}$ and a duct radius or perforation radius of 1 mm, this low frequency assumption is thus valid for frequencies much lower than 5~400 Hz (i.e. already a notable frequency range for human-audible acoustics). Another noticeable point of the low frequency value of the length correction is that models of frequency dependent length correction use the low frequency as their starting point (see e.g. \cite{Kar53,KG87}). \\[3mm]

In section 2, we recall some results of the early literature on the length correction, at low frequencies and for linear acoustic levels, for a baffle duct opening or a single perforation: $\varepsilon_0$.

In section 3, we discuss the expressions of the length correction for various duct discontinuities or perforation patterns, always in the low frequency range and for linear acoustic levels ($\varepsilon$). Ducts or perforations with various geometries are considered. The ducts considered are coaxial while the perforations are straight and centred with respect to a pattern. 
We propose a novel interpretation of previous works neglecting the mode-coupling at a duct discontinuity or at the surface of a perforated plate. This interpretation leads to a single expression of the length correction for all geometries of perforations or duct-discontinuities, except slit-like geometries which do not share the same topology.  It is shown that the length correction can be written as $\varepsilon=\varepsilon_0\Psi$ where $\Psi$ is a function depending on the perforation pattern or surface ratio at the discontinuity. \\

In section 4, the effect of the thickness, $h$, of the ducts or perforations is investigated: $\varepsilon_0$ is rewritten $\varepsilon_0(h)$. \\

It is worth noting that the expressions of length corrections reported hereafter are obtained for plane waves at normal incidence. However, J.-F. Allard \cite{All93} (pages 249 and 252) has observed, in the cases of facings with circular or square perforations in a square pattern, that "the dependence on the incidence angle of $\varepsilon$ can be neglected" and the expressions at normal incidence can be used at oblique incidence.  Experimental works for other configurations have led to the same conclusion (see e.g. \cite{ML00,LNM00}). \\

\section{Case of a single perforation or a baffled duct opening}
\label{sec.PhysicalAnalysis}

In volume II of \textit{The Theory of Sound} (\S 307 and Appendix A), first published in 1878, John William Strutt (third baron Rayleigh) \cite{Ray89} presented an analysis of the standing wave pattern created when 
an acoustic wave propagates in a cylindrical duct of thickness $h$ and radius $r$ which opens in a semi-infinite space (i.e. duct opening with an infinite flange or baffled duct opening).
In this analysis, limited to low frequencies, i.e. when $k_0 r \ll 1$,
J. W. Strutt has theoretically verified G. Wertheim hypothesis stating that the origin of the standing wave pattern is not located at the boundary between the duct and the open space but is translated by a "correction to length" $\varepsilon$ which enables acoustic radiation of the duct in this open space.

From computations assuming two axial velocity profiles $v$ for the flow, constant over the duct section (i.e. piston motion) or a polynomial function of the distance from the center of the duct to its wall, J. W. Strutt has established the length correction of the duct, or of the perforation, at low frequencies, $\varepsilon_0$, lies between
\begin{equation}
\varepsilon_{0}^{v \textrm{ constant}} = \frac{8r}{3 \pi} \simeq 0.85 r
\label{eq.varepsilon_constant_velocity_r}
\end{equation}
and
\begin{equation}
\varepsilon_{0}^{v \textrm{ polynomial},\ h\gg r}\simeq 0.82 r
\label{eq.varepsilon_polynomial_velocity_r}
\end{equation}
The first equation, which corresponds to the length correction for a plane piston excitation in a duct with a circular cross-section and with an infinite flange size at large wavelengths, is commonly re-written for a duct with a non-circular cross-section as:
\begin{equation}
\varepsilon_{0}^{v \textrm{ constant}} = \frac{8}{3 \pi^{3/2}} \sqrt{S_1}
\label{eq.varepsilon_constant_velocity_A0}
\end{equation}
where $S_1$ denotes the surface area of the duct cross-section. 

Equation (\ref{eq.varepsilon_polynomial_velocity_r}) is the length correction accounting for the fact that even at very low frequencies and without considering dissipation, the velocity profile in the duct, in the vicinity of the opening, is not constant with respect to the cross-section area of the duct. Indeed, one effect of the radiation, which can be interpreted as the presence of transverse evanescent modes, modify the velocity profile (see e.g. \cite{KG87}).

The condition $h\gg r$ in eq. (\ref{eq.varepsilon_polynomial_velocity_r}) is not explicitly written by J. W. Strutt. However he assumes no influence of the duct length on the length correction which was later shown to be correct for $h/r\gg 1$ (see section \ref{sec.InfluenceOfPerforationThickness}, figure \ref{fig.SigmoidFitFromKomkinEtAl2012} and e.g. \cite{KGTD89}).

When the thickness of the duct or the perforation $h$ tends to $0$ (the duct or the perforation is sometimes called a [thin] diaphragm in such a case), H. von Helmholtz, then J. W. Strutt, have concluded using different developments that the length correction, on one single side of the diaphragm with an infinite flange and at low frequencies is:
\begin{equation}
\varepsilon_{0}^{h\rightarrow 0}= \pi r /4 \simeq 0.79 r
\label{eq.varepsilon_contant_velocity_r_h_smaller_than_r}
\end{equation}

The small difference between the expressions of the length correction in eqs. 
(\ref{eq.varepsilon_polynomial_velocity_r}) and (\ref{eq.varepsilon_contant_velocity_r_h_smaller_than_r}) reflects the weak interaction between the evanescent modes inside and outside the diaphragm.

This small difference has also led
some authors, including J. W. Strutt, to consider the length correction is "independent, or nearly so, of the thickness of the duct". The dependence of the length correction with respect to the duct or perforation thickness will however be further studied in section \ref{sec.InfluenceOfPerforationThickness}.

As a reminder, eqs. (\ref{eq.varepsilon_constant_velocity_r}), (\ref{eq.varepsilon_polynomial_velocity_r}), (\ref{eq.varepsilon_constant_velocity_A0}) and (\ref{eq.varepsilon_contant_velocity_r_h_smaller_than_r}) are obtained at low frequencies for a duct or a single perforation in an infinite baffle so that the cross-section shape of the duct have a negligible influence on the radiation effect. Multiple perforations or finite discontinuities are discussed in the next section as well as the influence of their shapes.

\begin{figure}
\includegraphics[width=80mm]{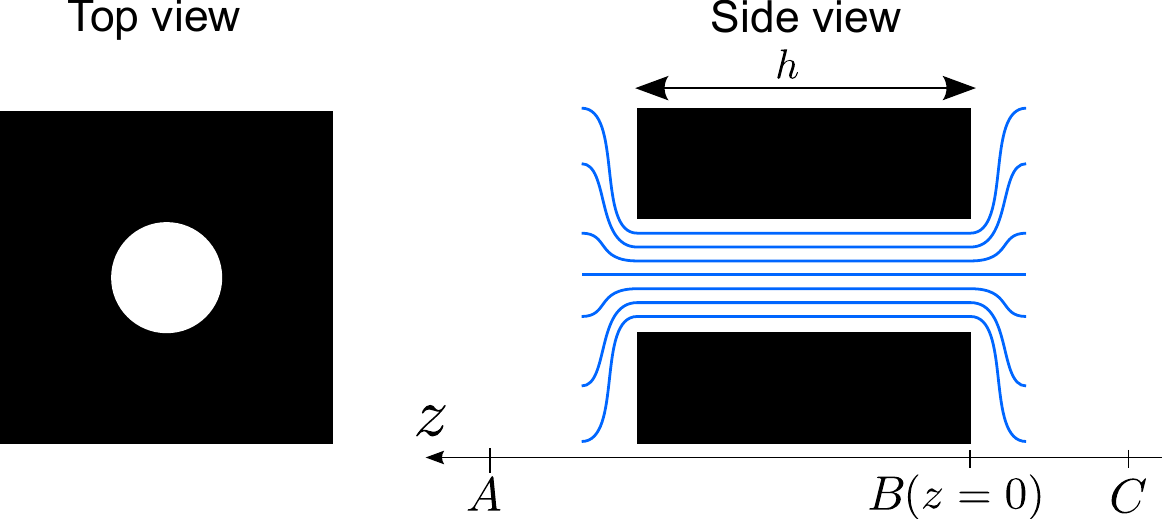} 
\caption{Schemes of a perforation or duct of thickness $h$ (side view and top view) with streamlines reported in the side view.}
\label{fig.PerforationTopAndSideViewsStreamlinesVisible}
\end{figure}

\section{Multiple perforations or finite duct discontinuities\label{sec.Deltal}}

\begin{figure}
\includegraphics[width=80mm]{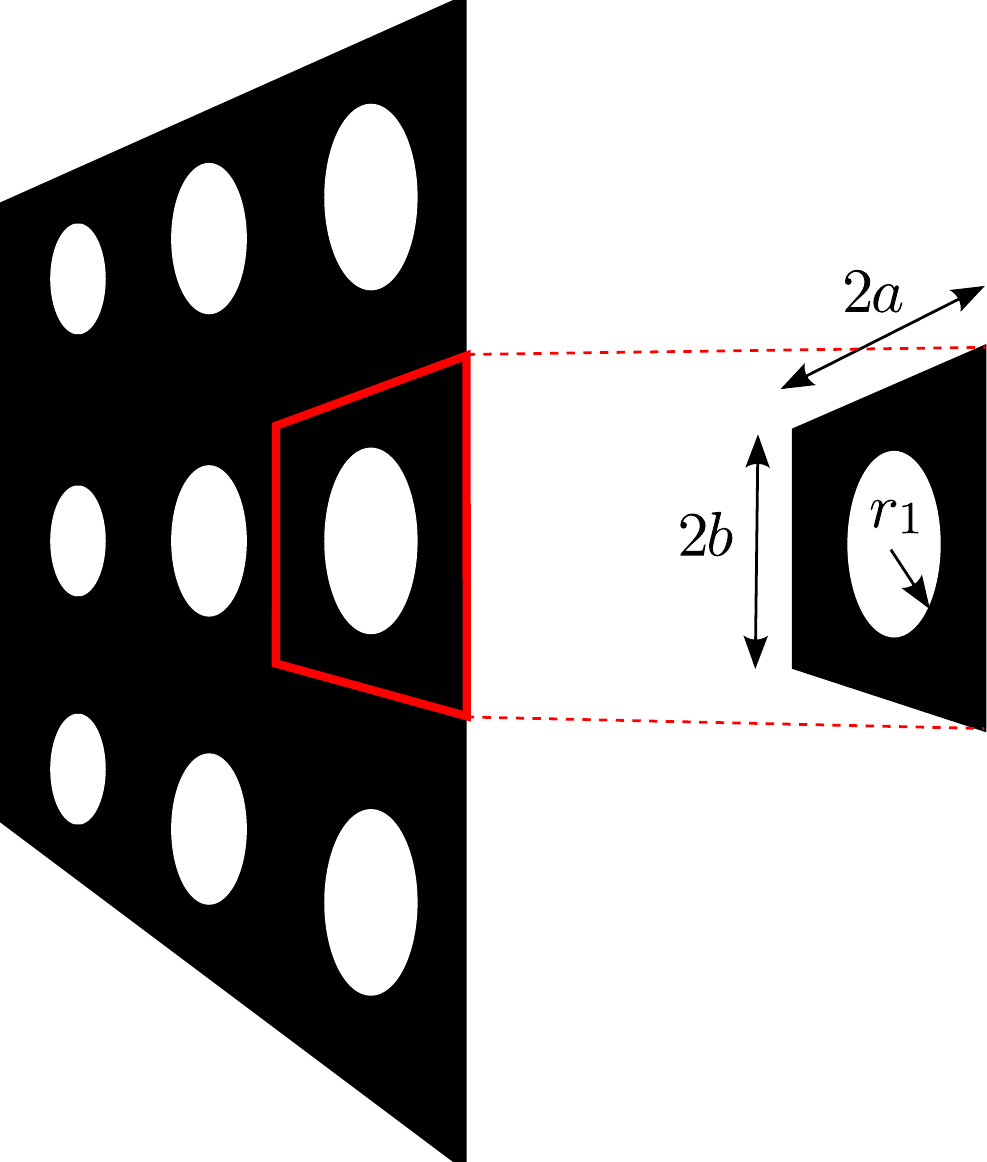} 
\caption{Scheme of a perforated structure composed by equidistant and identical circular perforations with a radius $r_1$ evenly distributed in a rectangular pattern of area $2a\times 2b$ (perspective views).}
\label{fig.PerforatedFacingToPattern}
\end{figure}

In the case of multiple straight perforations centered with respect to a pattern (see figure \ref{fig.PerforatedFacingToPattern}), the periodicity of the perforations (which can be associated to a symmetry) and the symmetry of the sound excitation (plane waves or any combination of plane waves, i.e. diffuse field), imply the linear acoustic problem, at large wavelengths, can be solved by studying only one pattern. In other words, the sound propagation in one pattern is considered to behave independently from its neighbors at least as the pattern dimensions are much smaller than the acoustic wavelength. Therefore, the study of the acoustic radiation for a perforated plate is, under such conditions, similar to the one for a duct discontinuity and the term "hole interaction" takes in this case a particular meaning that have misled numerous authors as the acoustic flow only passes through one perforation and is constrained by the neighboring perforations.

As an aside, it can be shown that the confinement of waves in a pattern is a valid approximation even for non-identical perforations randomly located such as non-woven textiles for which characteristic lengths, much smaller than the acoustic wavelength, denoting a periodicity of the perforations and a perforation dimension exist (see e.g. \cite{San80,ABG09}.) 
Furthermore, the reader is invited to refer to e.g. \cite{Ing53b,Ih93} for developments on non-centered perforations and e.g. \cite{LPBF+14,HBKA+13} for oblique, conical or tapered perforations. \\

In a theoretical study dated from 1941, V. A. Fok \cite{Fok41} proposed an expression, at low frequencies, for the length correction of a thin circular diaphragm of diameter $d$ in a duct with a circular diameter $D$ (denoted hereafter with subscript $\circ\!\!\subset\!\!\CIRCLE$). This expression is based on a Taylor series expansion of Bessel functions for the variable $\xi = d/D$ (which is equivalent to the square root of the perforation rate $\phi$ : $\phi = \pi d^2 / [\pi D^2]=d^2 / D^2$).

\begin{align}
\Psi^{\textrm{Fok}}_{\circ\subset\CIRCLE} &= 1+a_1\xi+a_2\xi^2+a_3\xi^3+... +a_{12}\xi^{12}\nonumber \\
\varepsilon^{\textrm{Fok}}_{\circ\subset\CIRCLE} &= \varepsilon_0 \Psi^{\textrm{Fok}}_{\circ\subset\CIRCLE} 
\label{eq.Fok}
\end{align}
with 
 \begin{align*}
 a_1&=-1.40925, &a_2&=0 \\
 a_3&=0.33818, &a_4&=0 \\
 a_5&=0.06793, &a_6&=-0.02287 \\
 a_7&=0.03015, &a_8&=-0.01641 \\
 a_9&=0.01729, &a_{10}&=-0.01248 \\
 a_{11}&=0.01205, &a_{12}&=-0.00985
\end{align*}
In this latter expression, and in the following ones, the length correction $\varepsilon$ is expressed as the product of the length correction for one perforation with an infinite baffle $\varepsilon_0$ and a function $\Psi$ taking into account the geometrical ratio $\xi$ (here equivalent to $\sqrt{\phi}$).

In the configuration studied by Fok, the expression of $\varepsilon_0$ was given by twice the expression from equation (\ref{eq.varepsilon_contant_velocity_r_h_smaller_than_r}): twice the length correction on each side of a thin diaphragm with an infinite flange at low frequencies.

At the end of his article, Fok suggests to use, "for practical purposes", only powers of 0, 1, 3 and 5 on $\xi$ and rounded values at 0.01 of their coefficients: 
\begin{equation}
\Psi^{\textrm{Fok}}_{\circ\subset\CIRCLE} \simeq 1 - 1.41\xi + 0.34\xi^3 + 0.07\xi^5
\label{eq.Psi_Fok_reducedOrders}
\end{equation} 
which "also gives the correct value for $\Psi^{\textrm{Fok}}_{\circ\subset\CIRCLE}$ (i.e. 0) when $\xi = 1$" (i.e when $d=D$).
In the same issue of the Proceedings of the USSR Academy of Sciences, V. S. Nesterov \cite{Nes41} presented an experimental study for the same configuration as Fok. Nesterov proposed a polynomial fit to his measurements based on Fok's results, i.e. only powers of 0, 1 and 3 on $\xi$ were used.

\begin{equation}
\Psi^{\textrm{Nesterov}}_{\circ\subset\CIRCLE} = 1-1.47\xi+0.47\xi^3 
\label{eq.Nesterov}
\end{equation}
The associated length correction is written:
\begin{equation}
\varepsilon^{\textrm{Nesterov}}_{\circ\subset\CIRCLE}= \varepsilon_0\Psi^{\textrm{Nesterov}}_{\circ\subset\CIRCLE}
\end{equation}

Nesterov gives a range of validity for his formula: $0 \leq \xi = \sqrt{\phi} \leq 0.9$ i.e, "the entire range of practical interest" to cite his words.

Fok and Nesterov works have been mainly disseminated outside Russia by their compatriot S. N. Rschevkin \cite{Rsc63} who published articles in French and German while his book, first written in Russian in 1960, was translated into English in 1963. This leads some authors to erroneously attribute Fok's or Nesterov's works to Rschevkin.\\

Obviously, the expressions of the length correction given by Fok and Nesterov ($\varepsilon^{\textrm{Fok}}_{\circ\subset\CIRCLE}$ and $\varepsilon^{\textrm{Nesterov}}_{\circ\subset\CIRCLE}$) give close results. For a low perforation rate ($d\ll D$), the expression of the length correction computed by H. von Helmholtz and J. W. Strutt for a diaphragm of radius $r$ within an infinite baffle is recovered. For a perforation rate approaching 1 ($d\rightarrow D$) the length correction tends to 0. \\[3mm]

\begin{figure}[h!]
\includegraphics[width=80mm]{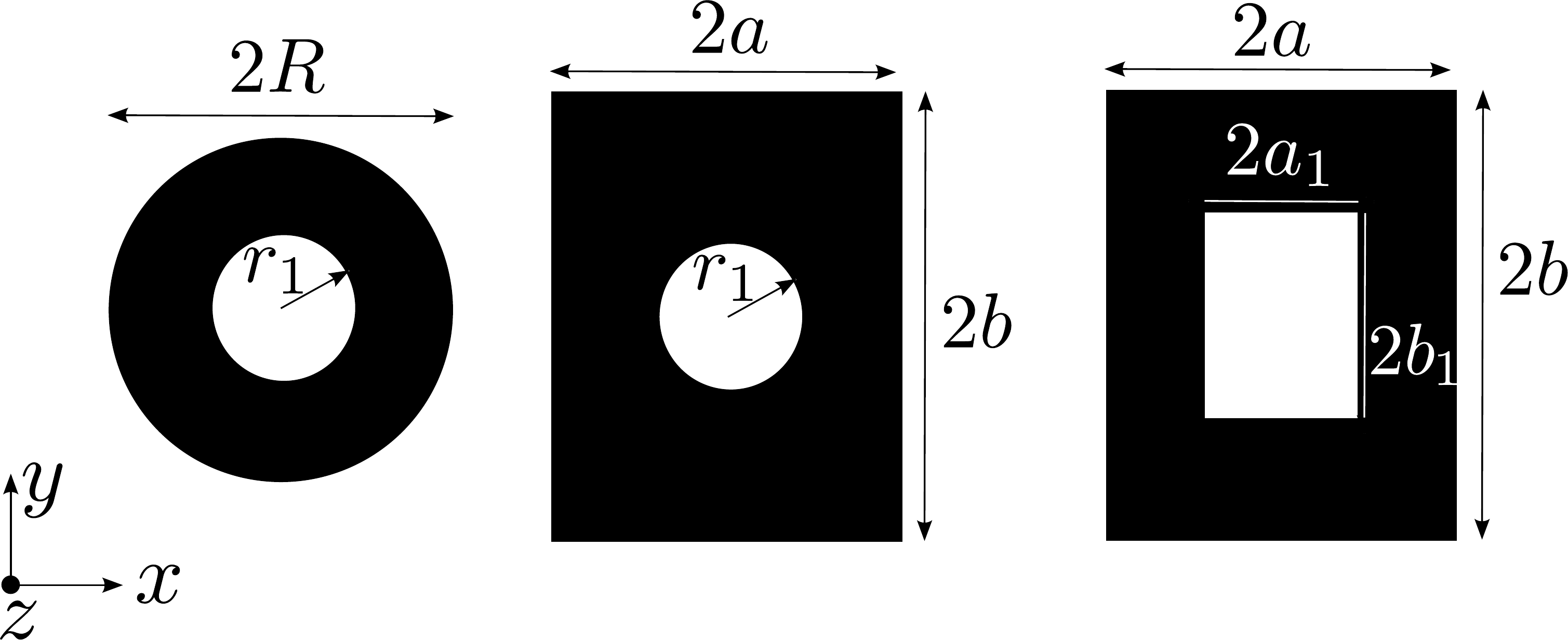} 
\caption{Schemes of the perforation geometries studied by U. Ingard in 1953 (top view)}
\label{fig.perfos}
\end{figure}

In a publication dedicated to acoustic resonators and published in 1948, U. Ingard \cite{Ing48} derived an expression of the length correction for a circular perforation in a circular pattern. This expression will be derived independently by F. C. Karal \cite{Kar53} in 1953. The same year U. Ingard extended his own work to additional geometries: rectangular perforation in a rectangular pattern, circular perforation in a rectangular pattern \cite{Ing53b} (see fig. \ref{fig.perfos}). These two additional geometries are respectively denoted hereafter with subscripts $\Box\!\!\subset\!\!\blacksquare$ and $\circ\!\subset\!\blacksquare$).

The expressions of the length corrections derived by Ingard (free from typos) appearing in his 1953's work, are reported in table \ref{table.ExpressionsOfLengthCorrections}. The complete derivations for these geometries are reported in a dedicated page on APMR website \cite{APMR}. Note that Ingard has assumed plane piston radiations in his work so that $\varepsilon_{0}$ is given by eq. (\ref{eq.varepsilon_constant_velocity_r}) for circular perforations or by eq. (\ref{eq.varepsilon_constant_velocity_A0}) for the rectangular perforation.

The expression of the length correction for the case of a circular perforation in a circular pattern, reported by U. Ingard and F. C. Karal, $\varepsilon_{\circ\subset\CIRCLE}$, is:

\begin{equation}
\varepsilon_{\circ\subset\CIRCLE} = 4R \sum_{m^*}\frac{J^2_1(k_m r_1)}{J^2_0(k_m R)} \frac{1}{(k_m R)^3}
\end{equation}

where the values of the discrete wavenumbers $k_m$ are the solutions of $J_1(k_m R)=0$ ($J_\alpha$ denotes the Bessel function of the first kind and of order $\alpha$).

\begin{figure}[h!]
\includegraphics[width=80mm]{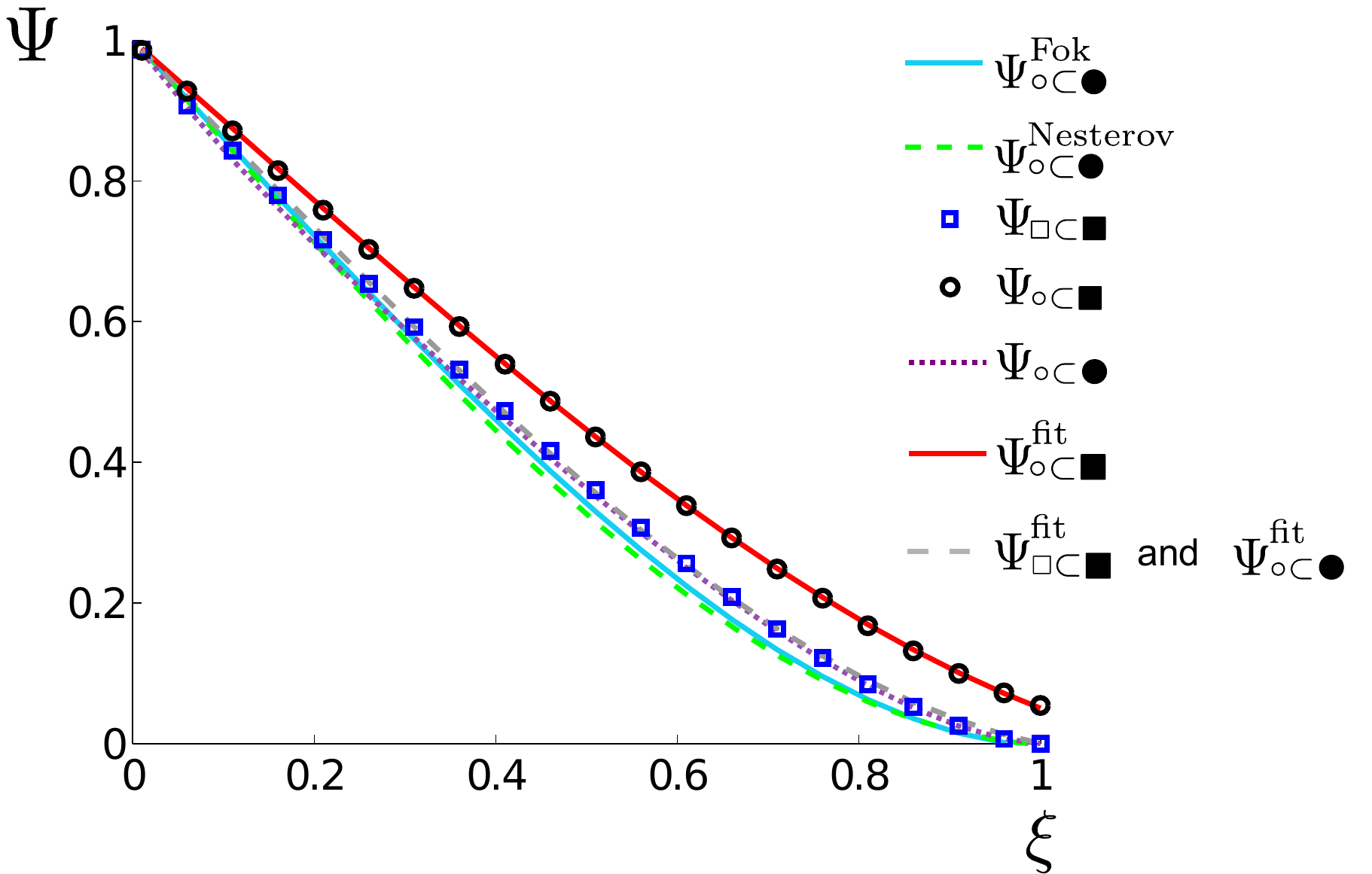}
\caption{Pattern functions $\Psi$ and their fits used in length corrections ($\varepsilon = \varepsilon_0\Psi$) for various perforation morphologies as a function of dimensionless variable $\xi$ (which corresponds to the square root of the perforation $\phi$ for cases $\circ\!\subset\!\CIRCLE$ and $\Box\!\subset\!\blacksquare$).}
\label{fig.LengthCorrectionsAsFunctionsOfXi_MyFits_merged}
\end{figure}

As it can be observed in figure \ref{fig.LengthCorrectionsAsFunctionsOfXi_MyFits_merged}, the expression of $\Psi_{\circ\subset\CIRCLE} = \varepsilon_{\circ\subset\CIRCLE}/\varepsilon_0$ derived by U. Ingard or F. C. Karal and computed in this article with more than 10$^4$ modes in the circular pattern gives close results compared to the expressions of V. Fok and to V. Nesterov functions. This observation confirms, once again, that the mode-coupling at the duct interface or at the surface of the perforation can be neglected in a first approximation.

In addition, the expressions of $\Psi_{\circ\subset\CIRCLE} = \varepsilon_{\circ\subset\CIRCLE}/\varepsilon_0$ and $\Psi_{\Box\subset\blacksquare} = \varepsilon_{\Box\subset\blacksquare}/\varepsilon_0$ for the specific case of a square perforation in a square pattern ($a=b$ and $a_1=b_1$) are so close to each others for all values of $\xi$, that, for the scale used in
figure \ref{fig.LengthCorrectionsAsFunctionsOfXi_MyFits_merged} 
they almost cannot be differentiated. This result was already pointed out by U. Ingard in 1953. 

To avoid dealing with time and numerical consuming expressions to compute length corrections, U. Ingard has proposed an approximation, eq. (\ref{eq.Ingard_Approx}), for two configurations which give close results at low perforation rates: $\Box\!\subset\!\blacksquare$ a square perforation in a square pattern and $\circ\!\subset\!\CIRCLE$ a circular perforation in a circular pattern.

\begin{equation}
\varepsilon^{\xi<0.4}_{\Box\subset\blacksquare} = \varepsilon_0 \left(1-1.25\xi\right) = \varepsilon_0 \Psi^{\xi<0.4}_{\circ\subset\CIRCLE}  \approx \varepsilon^{\xi<0.4}_{\circ\subset\CIRCLE} 
\label{eq.Ingard_Approx}
\end{equation}

where the definition of $\xi$ depends on the studied case. $\xi = a_1 / a \ (= \eta)$ for case $\Box\!\subset\!\blacksquare$. $\xi$ is the ratio of the perforation diameter to the pattern diameter for case $\circ\!\subset\!\CIRCLE$ (see previous paragraph concerning Fok's work: $\xi = d/D$). The domain of validity proposed by Ingard, probably obtained from a visual observation, is $\xi < 0.4$ which corresponds to a perforation rate $\phi < 0.16$.

To overcome the low perforation rate limitation, we propose a fit of Ingard's theoretical results for a square perforation in a square pattern valid for any value of $\xi$ in the range $[0 - 1]$ (polynomial fit up to order 3 with constrained points \{0;1\} and \{1;0\}): 

\begin{align}
\Psi_{\square\subset\blacksquare}^{\textrm{fit}} &=1-1.33\xi-0.07\xi^2+0.40\xi^3 = \Psi_{\circ\subset\CIRCLE}^{\textrm{fit}}  \label{eq.PsiFitAllGeometries}\\
\varepsilon_{\square\subset\blacksquare}^{\textrm{fit}} &= \varepsilon_0 \Psi_{\square\subset\blacksquare}^{\textrm{fit}}
\end{align}

This latest fit of the theoretical result by Ingard can also be used in the case of a circular perforation in a circular pattern as it has been shown the two configurations give very close results.\\

In the first edition of \textit{Propagation of sound in porous media}, J.-F. Allard \cite{All93} introduced a different approximation for the specific case of a circular perforation in a square pattern:
\begin{equation}
\varepsilon^{\xi < 0.4}_{\circ\subset\blacksquare} = \varepsilon_0 \left(1-1.14\xi\right) = \varepsilon_0 \Psi^{\xi < 0.4}_{\circ\subset\blacksquare}
\label{eq.Allard_1993_corrected}
\end{equation}

The same domain of validity: $\xi < 0.4$ (i.e. $\phi < 0.13$ where $\phi = \pi r_1^2/(2a)^2 = \pi \xi^2/4$) is defined by Allard for this approximation. Note that a typo appears in the two published editions of Allard's book and in various works based on them: $\sqrt{\phi}$ is used where one should read $\xi$.

In an article dedicated to the experimental characterization of perforated facings published in 2011, Jaouen \& B\'{e}cot \cite{JB11} introduced a function which is a fit up to order 3 of Ingard's theoretical results for a circular perforation in a square pattern ($\xi = \eta$) with a constrained point  \{0;1\}:

\begin{align}
\Psi_{\circ\subset\blacksquare}^{\textrm{fit}} &=1-1.13\xi-0.09\xi^2+0.27\xi^3 \\
\varepsilon_{\circ\subset\blacksquare}^{\textrm{fit}} &= \varepsilon_0 \*\Psi_{\circ\subset\blacksquare}^{\textrm{fit}}
\label{eq.Psi_JaouenBecot2011}
\end{align}

This expression is valid for any value of $\xi$ in the range $[0 - 1]$ as observed in figure \ref{fig.LengthCorrectionsAsFunctionsOfXi_MyFits_merged}.

\onecolumn
\thispagestyle{empty}
\begin{table}[!ht]\centering
\caption{Low frequency expressions of the length correction for different perforation geometries
The general expressions are those derived by U. Ingard \cite{Ing53b} corrected from typos. The value of $\varepsilon_0$ in Ingard's original work is given by eq (\ref{eq.varepsilon_constant_velocity_r}) or (\ref{eq.varepsilon_constant_velocity_A0}) as discussed in section \ref{sec.PhysicalAnalysis}. A more general expression of $\varepsilon_0$ is given in section \ref{sec.InfluenceOfPerforationThickness}. $mn^*$ denotes a summation excluding the mode ($m=0$, $n=0$), $m^*$ denotes a summation excluding the mode ($m=0$).}
\begin{tabular}{p{35mm}@{\extracolsep{12mm}}p{120mm}}
\hline
\hline\\[-3mm]
Geometry & Expression of length correction\\
\hline\\
Rectangular perforation in a rectangular pattern & $\varepsilon_{\Box\subset\blacksquare} = \displaystyle\frac{4a_1 b_1}{\pi a b}\sum_{mn^*}\nu_{mn}\frac{\Big[\displaystyle\frac{\sin(\pi m a_1/a)}{\pi m a_1/a} \frac{\displaystyle\sin(\pi n b_1/b)}{\pi n b_1/b}\Big]^2}{\sqrt{\left(\displaystyle\frac{m}{a}\right)^2+\left(\displaystyle\frac{n}{b}\right)^2}}$ \\[2mm]
 & $\nu_{0n} = \nu_{m0} = 1/2 \qquad \nu_{mn} = 1$ \\[2mm]
  & $\xi = a_1/a$ \hspace{10mm} $\eta = b_1/b$ \\[2mm]
\hline\\
 Circular perforation\hspace{5mm} in a rectangular pattern & $\varepsilon_{\circ\subset\blacksquare}=\displaystyle\frac{4\pi}{ab} \sum_{mn*} \nu_{mn} \frac{J_1^2\left(r_1\sqrt{\displaystyle\left(\frac{\pi m}{a}\right)^2+\left(\frac{\pi n}{b}\right)^2}\right)}{\left[\displaystyle\left(\frac{\pi m}{a}\right)^2+\left(\frac{\pi n}{b}\right)^2\right]^{3/2}}$ \\[2mm]
  & $\nu_{0n} = \nu_{m0} = 1/2 \qquad \nu_{mn} = 1$ \\[2mm]
 & $\xi = r_1/a$ \hspace{10mm} $\eta = r_1/b$ \\[2mm]
 \hline\\[2mm]
Circular perforation\hspace{5mm} in a circular pattern & $\varepsilon_{\circ\!\subset\!\CIRCLE} = 4R \displaystyle\sum_{m^*} \displaystyle\frac{J^2_1(k_m r_1)}{J^2_0(k_m R)} \displaystyle\frac{1}{(k_m R)^3}$ \\[2mm]
 &  $\xi = r_1/R$ and $k_m$ are the solutions of $J_1(k_m R)=0$  \\[2mm]
  \hline\\
  Approx. all geometries
  & $\varepsilon \simeq \varepsilon_0  \Big(  1-1.33\sqrt{\phi}-0.07\sqrt{\phi}^2+0.40\sqrt{\phi}^3 \Big)$  \\[2mm]
   (except slit-like) 
   & where $\phi$ is the perforation ratio or the ratio of the duct cross-section areas at a discontinuity.  \\
   & Formulas by Fok (\ref{eq.Psi_Fok_reducedOrders}), Nesterov (\ref{eq.Nesterov}) or Jaouen \& Bécot (\ref{eq.Psi_JaouenBecot2011}) can also be applied if $\xi$ variables are expressed as functions of $\sqrt{\phi}$.\\[2mm]
\hline
\hline
\end{tabular}
\label{table.ExpressionsOfLengthCorrections}
\end{table}%
\twocolumn

\begin{figure}[h!]
\includegraphics[width=80mm]{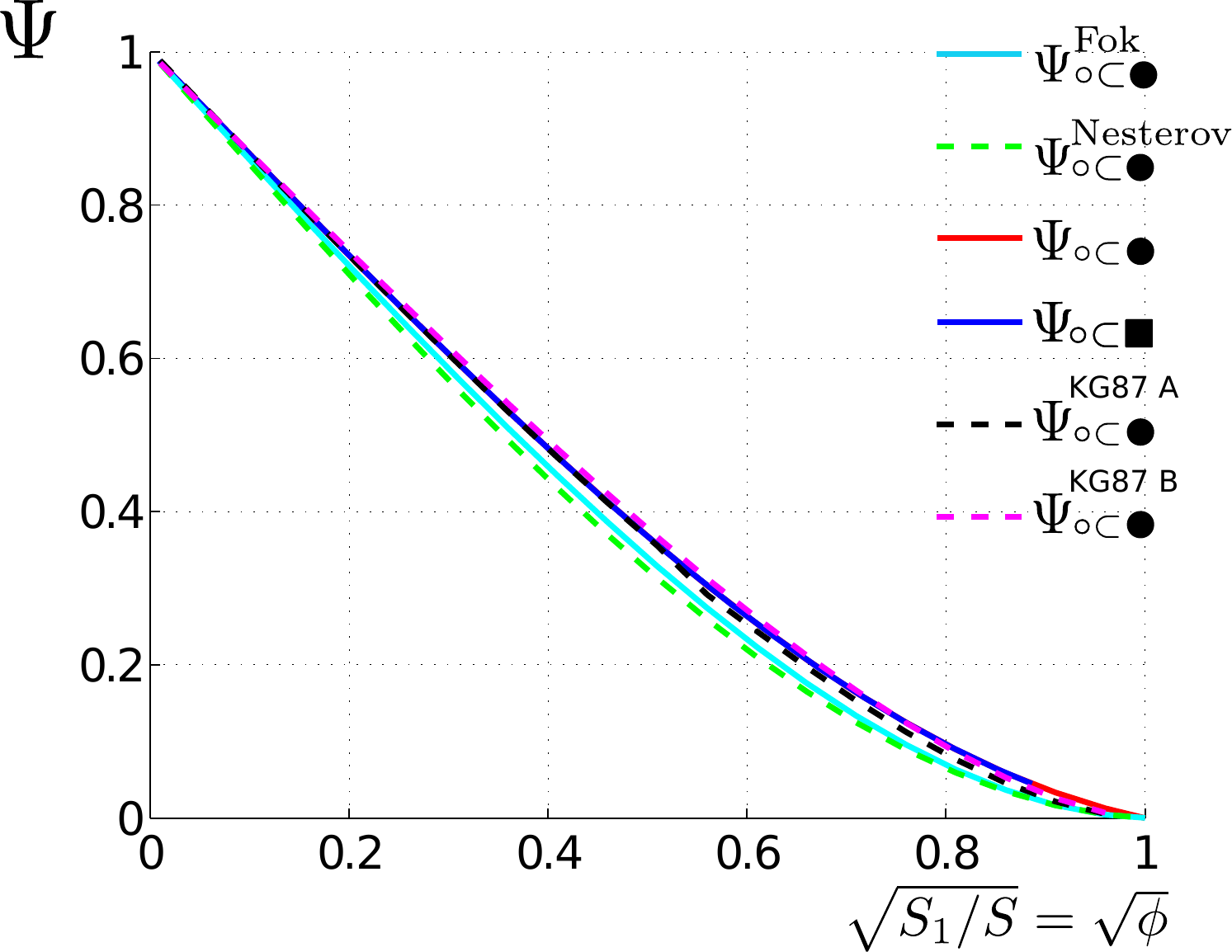}
\caption{Pattern functions $\Psi$ used in length corrections ($\varepsilon = \varepsilon_0\Psi$) for various perforation morphologies (circular perforation in circular or square pattern, square perforation in square pattern) as a function of dimensionless variable $\sqrt{\phi}$.}
\label{fig.LengthCorrections_sqrtphi}
\end{figure}

When plotting the expressions reported above as a function of the square root of the perforation cross-section area ($S_1$) over the pattern cross-section area ($S$): $\sqrt{S_1/S}=\sqrt{\phi}$, it appears they give almost the same result (see fig. \ref{fig.LengthCorrections_sqrtphi}).
Some additional approximations are reported in figure \ref{fig.LengthCorrections_sqrtphi}: the low frequency approximation for a plane piston excitation published by Kergomard \& Garcia in 1987 \cite{KG87} ($\Psi_{\circ\subset\CIRCLE}^{\textbf{KG87A}}$), the long duct approximation, again at low frequencies, also published by Kergomard \& Garcia in 1987 ($\Psi_{\circ\subset\CIRCLE}^{\textbf{KG87B}}$).

The maximum deviation, 0.03, observed between these curves is obtained with $\Psi^{\textrm{Fok}}_{\circ\subset\CIRCLE}$ and $\Psi_{\circ\subset\CIRCLE}^{\textbf{KG87B}}$ around $\sqrt{\phi}$ = 0.6.
The mean distance between these two curves, as computed using eq. (\ref{eq.MeanDistanceBetween2Curves}), is less than 1\%: 0.007

From the analysis of the results obtained above, two conclusions can be deduced.
\begin{itemize}
\item[$\bullet$] First, comparing $\Psi$ functions obtained for various duct or perforation thicknesses, the mode coupling at a duct discontinuity or in the vicinities of a perforation appears to be weak for low frequencies. $\Psi$ functions do not strongly depend on the thickness of the duct or diaphragm and do not strongly depend on the values of $\varepsilon_0$ ($\pi/4$ for a diaphragm with a thickness which tends to $0$, 0.82 for long duct, $8\sqrt{S_1}/(3 \pi^{3/2})$ for a piston motion). The effect of the tube or perforation thickness is further discussed in the following section.
\item[$\bullet$] Second, comparing $\Psi$ functions for various geometries, it appears these $\Psi$ functions do not strongly depend on the shape of the perforation or the duct discontinuity. Indeed, they can be considered as functions depending only on the perforation ratio or the surface ratio at the duct discontinuity.
\end{itemize}

\begin{figure}[h!]
\includegraphics[width=80mm]{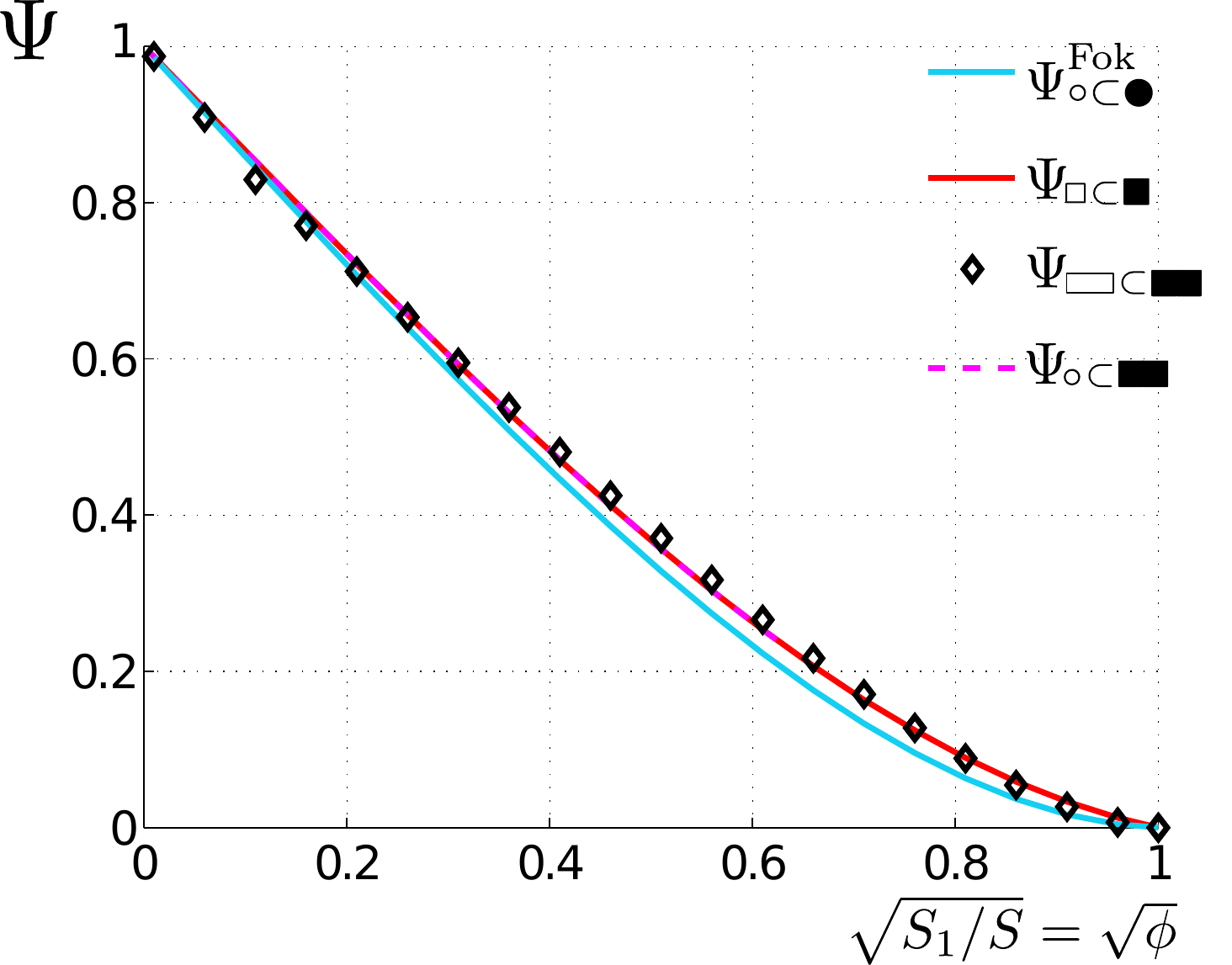}
\caption{Pattern functions $\Psi$ used in length corrections ($\varepsilon = \varepsilon_0\Psi$) for various perforation morphologies (circular perforation in circular pattern, square perforation in square pattern, circular or rectangular perforation in rectangular pattern) as a function of dimensionless variable $\sqrt{\phi}$.}
\label{fig.RectangularPerforationAsAFunctionOfSqrtPhi}
\end{figure}

These 2 points confirm the assumption of a length perforation written, in the first order approximation, as $\varepsilon = \varepsilon_0 \Psi$.
They also imply that a single formula can be used to calculate $\Psi$ and thus $\varepsilon$ for any 2D geometry:
\begin{eqnarray}
\Psi \simeq \Big(  1-1.33\sqrt{\phi}-0.07\sqrt{\phi}^2+0.40\sqrt{\phi}^3 \Big) 
\label{eq.Psi_JaouenChevillotte2018} 
\end{eqnarray}
To further illustrate the fact that a single $\Psi$ function (e.g. $\Psi_{\Box\subset\blacksquare}$ or $\Psi_{\Box\subset\blacksquare}^{\textrm{fit}}$...) can be used to approximate function $\Psi$ for a rectangular duct discontinuity (or rectangular perforation pattern), 
figure \ref{fig.RectangularPerforationAsAFunctionOfSqrtPhi} compares 4 $\Psi$ functions.
The first function is the one by Fok. The second is the one by Ingard for the square case. The third one ($\Psi_{\sqsubset\!\sqsupset\subset\blacksquare\!\blacksquare}$) is the one for the rectangular case with $a_1/b_1 = 2$ (and $a/b = 2$ so that $\xi=\eta$) and the fourth one ($\Psi_{\circ\subset\blacksquare\!\blacksquare}$), a circular duct discontinuity or perforation with the same surface area than $\Psi_{\sqsubset\!\sqsupset\subset\blacksquare\!\blacksquare}$.
The mean distance between the farthest curves: $\Psi^{\textrm{Fok}}_{\circ\subset\CIRCLE}$ and $\Psi_{\sqsubset\!\sqsupset\subset\blacksquare\!\blacksquare}$, as computed using eq. (\ref{eq.MeanDistanceBetween2Curves}), is 
0.016

The figure \ref{fig.RectangularPerforations_ParamStudy.pdf} reports the mean distances, as computed using eq. (\ref{eq.MeanDistanceBetween2Curves}), between $\Psi_{\sqsubset\!\sqsupset\subset\blacksquare\!\blacksquare}$ functions and the function $\Psi$ computed as per eq. (\ref{eq.Psi_JaouenChevillotte2018}) 
 for various values of the ratios $a_1/b_1$ and $\xi/\eta$.

\begin{equation}
\textrm{mean\ distance} =  \sqrt{ \frac{1}{N} \sum_N  \Big( \Psi_{1,i} - \Psi_{2,i}  \Big)^2 }\label{eq.MeanDistanceBetween2Curves}
\end{equation}

Only physical geometries are considered when comparing 2 curves i.e. points for which one of the 2 functions $\Psi_{1,i}$ or $\Psi_{2,i}$ corresponds to $a < a_1$ or $b < b_1$ are not considered.


\begin{figure}[h!]
\includegraphics[width=80mm]{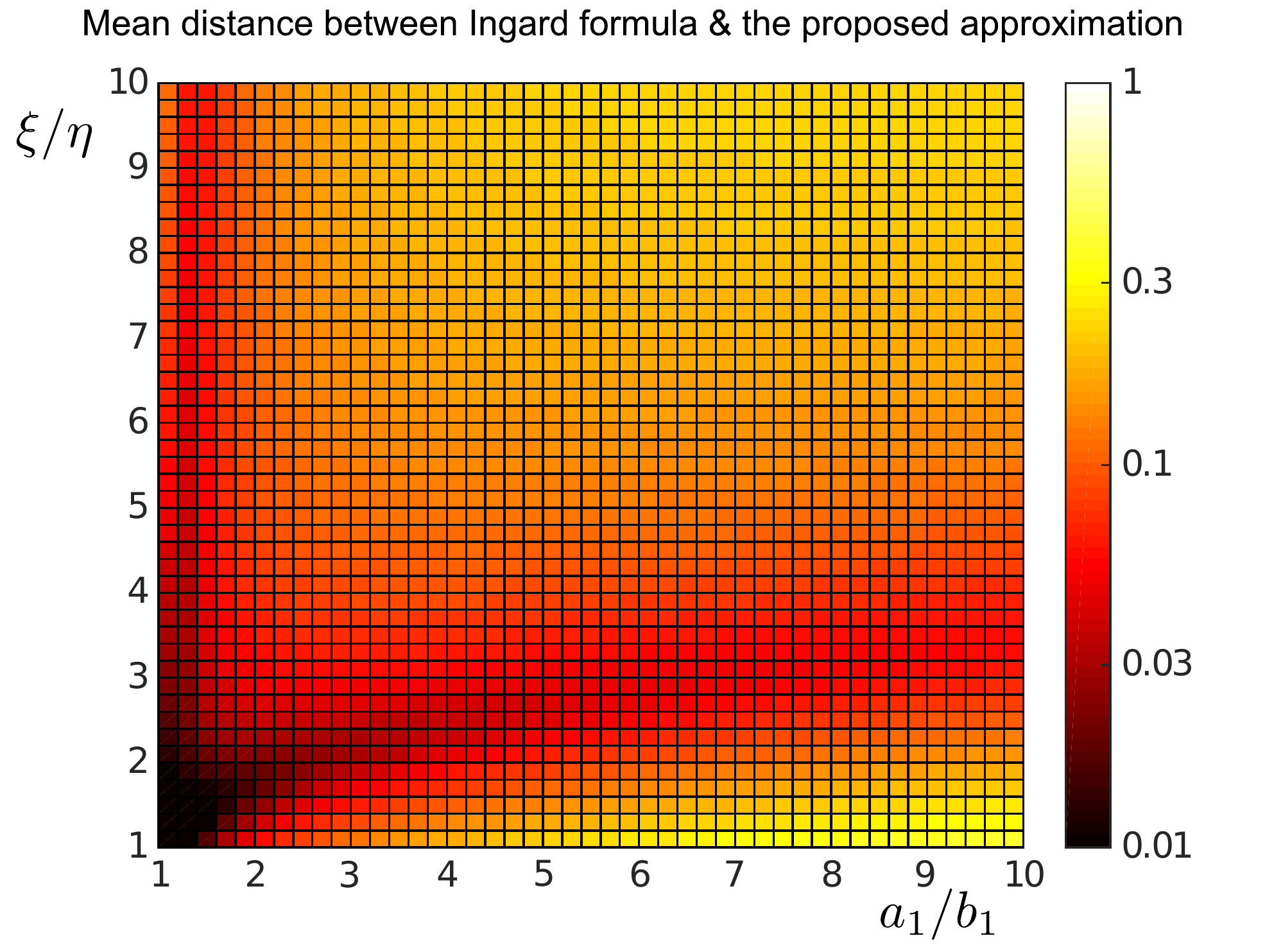}
\caption{Mean distances between $\Psi$ functions computed following Ingard's expression for rectangular perforation in a rectangular pattern and $\Psi$ function given in eq. (\ref{eq.Psi_JaouenChevillotte2018}) for various values of the ratios $a_1/b_1$ and $\xi/\eta$. For the sake of clarity, only positive values for the ratios are reported on this figure.}
\label{fig.RectangularPerforations_ParamStudy}
\end{figure}

In figure \ref{fig.RectangularPerforations_ParamStudy}, it appears the proposed approximation for the $\Psi$ function given in eq. (\ref{eq.Psi_JaouenChevillotte2018}) gives acceptable results  (with mean distances between Ingard's expression and eq. (\ref{eq.Psi_JaouenChevillotte2018}) lower or equal to 0.1) as long as the studied geometry does not tend to a slit-like geometry, i.e. $a_1/b_1$ or $\xi/\eta$ both not much smaller or much larger than 1. The best approximation is obtained for $a_1 = b_1$ or $\xi = \eta$ with deviation between the 2 curves equal or less than 0.01\ .

\begin{figure}[h!]
\includegraphics[width=80mm]{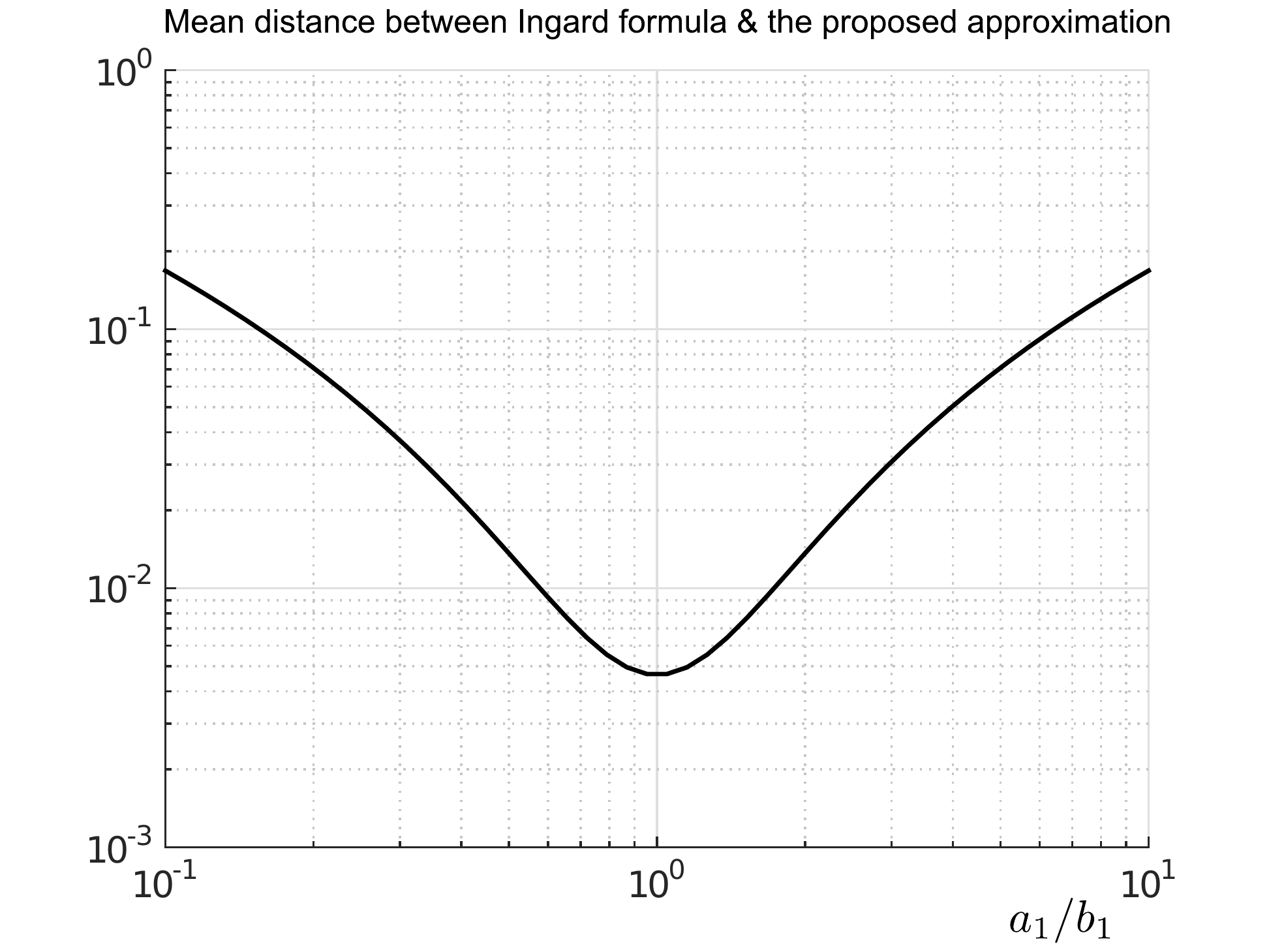}
\caption{Mean distances between $\Psi$ functions computed following Ingard's expression for rectangular perforation in a rectangular pattern and $\Psi$ function given in eq. (\ref{eq.Psi_JaouenChevillotte2018}) for different values of the ratios $a_1/b_1$ (with $\xi=\eta=0.5$).} 
\label{fig.ParamStudy_SurA1B1EtXiEta}
\end{figure}

Figure \ref{fig.ParamStudy_SurA1B1EtXiEta} shows detailed results for various values of the ratios $a_1/b_1$ ranging from 10$^{-1}$ to 10$^{1}$ while $\xi=\eta=0.5$. As expected, $a_1/b_1=1$ is a symmetry line for these results.

\section{Influence of the thickness \label{sec.InfluenceOfPerforationThickness}}

Following our hypothesis of weak coupling between the modes in ducts at a discontinuity or at the vicinities of  a perforation, we propose to account for the effect of the thickness, $h$, of the ducts or perforations by defining $\varepsilon_0$ as a function of $h$.

This thickness influence can be considered as a refinement of eq. (\ref{eq.Psi_JaouenChevillotte2018}) as the influence of the thickness on the length correction is much smaller than the influence of the pattern, $\Psi$, when $\Psi$ is not close to 1. \\

The results obtained, for circular geometries, by Kergomard et al. 1989 \cite{KGTD89} (see section 6.2 and figures 11a-c) show the non-dimensional ratio: length correction over perforation radius appears as a sigmoid function of the non-dimensional ratio: thickness of perforation over perforation radius in log-scale. 
The two asymptotic values are the length correction factor (on one side) for a diaphragm of thickness $h$ which tends to 0: $\varepsilon_0^{h\rightarrow 0}/r = \pi/4$ ($\simeq 0.79$) and the length correction factor for an infinite duct: $\varepsilon_0^{h/r\gg 1}/r \simeq 0.82$\ . 
Function $\varepsilon_0(h)/ r$ thus writes as:
\begin{equation}
\varepsilon_0(h)/ r = \frac{\displaystyle (\varepsilon_0^{h\rightarrow 0}/r) e^{d\cdot\log{(c)}}+ (\varepsilon_0^{h/r\gg 1}/r) e^{d\cdot \log{(h/r)}}}{e^{d\cdot \log{(c)}}+e^{d\cdot \log{(h/r)}}}
\label{eq.SigmoidFit}
\end{equation}

$c$ and $d$ are parameters defining the position of the inflection point and the slope of the sigmoid.

\begin{figure}[h!]
\includegraphics[width=80mm]{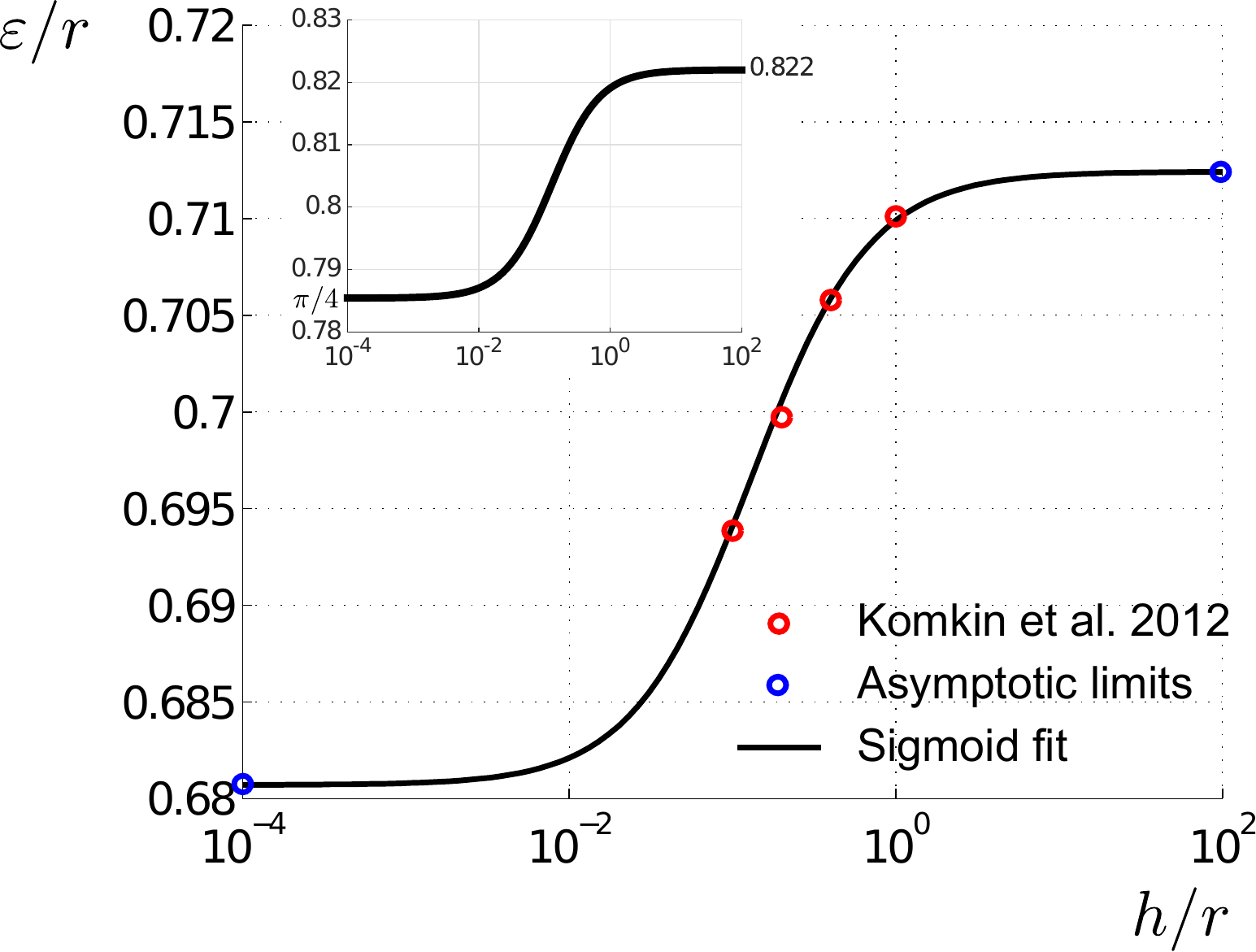}
\caption[dummy]{$\varepsilon/r = \varepsilon_0(h)/r \times \Psi_{\circ\subset\CIRCLE}(\xi=0.1)$ as a function of the dimensionless variable $h/r$ (with a logarithmic scale). The asymptotic limits are $\pi/4\times\Psi(0.1) \simeq 0.681$ for $h/r \rightarrow 0$ and $0.822\times\Psi(0.1) \simeq 0.712$ for $h/r \gg 1$. Inset figure: particular case when $\Psi=1$ ($\varepsilon/r = \varepsilon_0(h)/r$) with asymptotic limits $\pi/4$ and $0.822$.} 
\label{fig.SigmoidFitFromKomkinEtAl2012}
\end{figure}

An analytical inversion of the sigmoid function leading to the smallest mean square deviation with data from Komkin et al. \cite{KMY12} (mean values obtained for $\xi$ = 0.100, $\xi$ = 0.125 and $\xi$ = 0.200) is obtained with:
$c = 0.13$ and $d = 1.2$.
$\varepsilon_0(h)/ r$, with parameters reported above, as a function of $h/r$ (with a logarithmic scale) is plotted in figure \ref{fig.SigmoidFitFromKomkinEtAl2012}. 

This function is obviously coherent with the data from Komkin et al. 2012 (who were seeking for a linear dependence of $\varepsilon(h)$ as a function of $h/r$ which limits the validity of their interpretation). The sigmoid function is also coherent with the results published by Kergomard et al. 1989 \cite{KGTD89} with a small deviation for case $c$ ($\xi = 0.9$) difficult to quantify due to the ordinate scale used in Kergomard et al. 1989.

From the results of the previous section, this sigmoid correction factor can be used for non-cylindrical geometries by using their hydraulic radius as $r$.

\section{Conclusion}

In this work, we have proposed a single approximate expression for the length correction at low frequencies $\varepsilon$ which can be applied to all geometries of perforations or duct discontinuities, except slit-like geometries. This expression, obtained by neglecting the effect of mode coupling at a duct discontinuity or at the surface of a perforated plate, writes $\varepsilon = \varepsilon_0(h)\Psi$. $\varepsilon_0$ is the length correction for an open duct in an infinite baffle or the length correction on one side of a perforated plate with a single perforation. The expression of $\varepsilon_0$ is given in equation (\ref{eq.SigmoidFit}). $h$ is the thickness of the duct or perforated plate. $\Psi$ is a function depending on the perforation pattern or surface ratio at the discontinuity. $\Psi$ can be expressed using Ingard's 1948 work, Karal's 1953 work or by an order-3 polynomial fit with two constrained points (\{0;1\} and \{1;0\}) of these the two first similar functions (see eq. (\ref{eq.Psi_JaouenChevillotte2018}))

From the conclusions of the present work, a first approximation of the length corrections in the case of a tapered or conical perforation, not considered in the above study, consists in calculating two length corrections based on the perforation rate on each side (neglecting the weak influence of the perforation thickness).
This perspective point requires additional work to be validated. \\

A second perspective to this work is to compare frequency-dependent expressions of the length correction, currently studied independently with respect to the velocity profiles inside the ducts or perforations, while  
based on the results at the large wavelength limit (i.e. the limit studied in this work).

\section{Acknowledgment}
The authors acknowledge Jean Kergomard for the fruitful discussions they had during the review process of this work.


\end{document}